# Free-space propagation of spatio-temporal optical vortices (STOVs)

S. W. Hancock, S. Zahedpour, A. Goffin, and H. M. Milchberg

*Institute for Research in Electronics and Applied Physics, University of Maryland, College Park, Maryland 20742, USA*

**Spatio-temporal optical vortices (STOVs) are a new type of optical orbital angular momentum (OAM) with optical phase circulation in space-time. In prior work [N. Jhajj *et al.*, Phys. Rev X 6, 031037 (2016)], we demonstrated that a STOV is a universal structure emerging from the arrest of self-focusing collapse leading to nonlinear self-guiding in material media. Here, we demonstrate linear generation and propagation in free space of STOV-carrying pulses. Our measurements and simulations demonstrate STOV mediation of space-time energy flow within the pulse and conservation of OAM in space-time. Single-shot amplitude and phase images of STOVs are taken using a new diagnostic, transient grating single-shot supercontinuum spectral interferometry (TG-SSSI).**

Optical vortices are electromagnetic structures characterized by a rotational flow of energy density around a phase singularity, comprised of a null in the field amplitude and a discontinuity in azimuthal phase. In the most common type of optical vortex in a laser beam, the azimuthal phase circulation resides in spatial dimensions transverse to the propagation direction. An example is the well-known orbital angular momentum (OAM) modes [1], typified by Bessel-Gauss ($BG_l$) or Laguerre-Gaussian ($LG_{pl}$) modes with nonzero azimuthal index $l$. OAM beams have been used in optical trapping [2] and super-resolution microscopy [3], with proposed applications such as turbulence-resilient free space communications [4, 5] and quantum key distribution [6]. Optical vortex formation is ubiquitously observed in the speckle pattern of randomly scattered coherent light [7]. We note that all of these standard OAM vortices can, in principle, be supported by monochromatic beams and hence are fundamentally CW phenomena. Standard OAM vortices embedded in short pulse beams [8-10], necessarily polychromatic, have also been experimentally and theoretically studied, while theoretical work has explored polychromatic vortices that can exist in space-time [11].

Recently [12], we reported on the experimental discovery and analysis of the spatio-temporal optical vortex (STOV), whose phase winding resides in the spatio-temporal domain. Toroidal STOVs were found to be a universal electromagnetic structure that naturally emerges from arrested self-focusing collapse of short pulses, which occurs, for example, in femtosecond filamentation in air [13] or relativistic self-guiding in laser wakefield accelerators [14]. As this vortex is supported on the envelope of a short pulse, its description is necessarily polychromatic. For femtosecond filamentation in air, a pulse with no initial vorticity collapses and generates plasma at beam centre. The ultrafast onset of plasma provides sufficient transient phase shear to spawn two toroidal spatio-temporal vortex rings of charge $l = -1$ and $l = +1$ that wrap around the pulse. In air, the delayed rotational response of $N_2$ and $O_2$ [15] provides additional transient phase shear, generating additional $l = \pm 1$ ring STOVs on the trailing edge of the pulse [12]. After some propagation distance and STOV-STOV dynamics, the self-guided pulse is accompanied by the $l = +1$ vortex, which governs the intra-pulse energy flow supporting self-guiding [12].

The requirement of transient phase shear for such nonlinearly generated STOVs suggested that phase shear linearly applied in the spatio-spectral domain could also lead to STOVs, and use of a zero dispersion ($4f$) pulse shaper and phase masks have been proposed [16] and demonstrated [17] for this purpose. In this paper, we use such a $4f$ pulse shaper to impose STOVs on Gaussian pulses and record single-shot in-flight phase and amplitude images of these structures using a new diagnostic developed for this purpose. The structures generated are "line-STOVs" as described in [12, 16]; the phase circulates around a straight axis normal to the spatio-temporal plane. An electric field component of a simple $|l|^{th}$ order line-STOV–carrying pulse at position $z$ along the propagation axis can be described as

$$E(\mathbf{r}_\perp, z, \tau) = a(\tau/\tau_s + i \, \text{sgn}(l) \, x/x_s)^{|l|} E_0(\mathbf{r}_\perp, z, \tau) = A(x,\tau) e^{il\Phi_{s-t}} E_0(\mathbf{r}_\perp, z, \tau), \quad (1)$$

where $\mathbf{r}_\perp = (x, y)$, $\tau = t - z/v_g$ is a time coordinate local to the pulse, $v_g$ is the group velocity, $\tau_s$ and $x_s$ are temporal and spatial scale widths of the STOV, $\Phi_{s-t}(x, \tau)$ is the space-time phase circulation in $x - \tau$ space, $l = \pm 1, \pm 2, \ldots$, $A(x, \tau) = a((\tau/\tau_s)^2 + (x/x_s)^2)^{|l|/2}$, $a = \sqrt{2}((x_0/x_s)^2 + (\tau_0/\tau_s)^2)^{-1/2}$ for $l = \pm 1$, and $E_0$ is the STOV-free near-Gaussian pulse input to the shaper, where $x_0$ and $\tau_0$ are its spatial and temporal widths. Here $a$ is a normalization factor ensuring that pulse energy is conserved through the shaper: $\int d^2\mathbf{r}_\perp d\tau |E|^2 = \int d^2\mathbf{r}_\perp d\tau |E_0|^2$.

In order to confirm the presence of an ultrafast STOV-carrying pulse, both the phase and the amplitude of the electric field envelope must be measured, preferably in a single shot. In work by another group, amplitude and phase have been retrieved from femtosecond pulses undergoing filamentation using a multishot scanning technique in combination with an iterative algorithm [18]. In our group, we have used single shot supercontinuum spectral interferometry (SSSI) [19, 20] to measure the space- and time- resolved envelope (but not phase) of ultrafast laser pulses from the near-UV to the long wave infrared [21, 22]. It is worth first briefly reviewing SSSI. Figure 1 shows the three beams employed in SSSI: a pump pulse $E_S$ (the optical signal of interest, here a STOV) and twin supercontinuum (SC) reference and probe pulses $E_{ref}$ and $E_{pr}$. The reference and probe SC pulses are generated upstream of Fig. 1 in a 2 atm $SF_6$ cell followed by a Michelson interferometer (not shown). The transient amplitude of $E_S$ is measured via the phase modulation it induces in a spatially and temporally overlapped chirped supercontinuum (SC) probe pulse $E_{pr}$ in a thin instantaneous Kerr "witness plate" such as the thin fused silica window used here. The resulting spatio-spectral phase shift $\Delta\varphi(x, \omega)$ imposed on the probe is extracted from interfering $E_{pr}^{out} \sim \chi^{(3)} E_S E_S^* E_{pr}^{in}$ with $E_{ref}$ in an imaging spectrometer. Here, $E_{pr}^{in}$ and $E_{pr}^{out}$ are the probe fields entering and exiting the fused silica witness plate, $\chi^{(3)}$ is the fused silica nonlinear susceptibility, and $x$ is position within a 1D transverse spatial slice through the pump pulse at the witness plate (axes shown in Fig. 1). Fourier analysis of the extracted $\Delta\varphi(x, \omega)$ [19] then determines the spatio-temporal phase shift $\Delta\phi(x, \tau) \propto I_S(x, \tau)$, where $I_S(x, \tau)$ is the 1D space + time pump intensity envelope.

For measurements of STOVs, in which space-time phase circulation is the key feature, ordinary SSSI is insufficient. To measure space-time resolved amplitude *and* phase in a single shot, we have developed a new diagnostic, transient grating SSSI (TG-SSSI). In TG-SSSI a weak auxiliary probe pulse $\mathcal{E}_i$ (same central wavelength of the pump pulse and spectrally filtered by a 2nm bandpass filter, see Fig. 1) is interfered with the pump to form a transient spatial interference grating in the witness plate. Here $\mathcal{E}_i$ is crossed with $E_S$ at an angle $\theta = 6°$. The transient grating is now the structure probed by SSSI, with the output probe pulse becoming $E_{pr}^{out} \sim \chi^{(3)} E_S \mathcal{E}_i^* E_{pr}^{in}$. As before, $\Delta\varphi(x, \omega)$ is extracted from the interference of $E_{pr}^{out}$ and $E_{ref}$ in the imaging spectrometer, leading to $\Delta\phi(x, \tau)$. Now, however, $\Delta\phi(x, \tau)$ is encoded with the pump envelope modulated by the time-dependent spatial interference pattern (transient grating): $\Delta\phi(x, \tau) \propto I_S(x, \tau) f(x, \tau)$, where $k$ is the pump wavenumber, $f(x, \tau) = \cos(kx \sin\theta + \Delta\Phi(x, \tau))$ is the transient grating, and $\Delta\Phi(x,\tau)$ is the spatiotemporal phase of $E_S$. In the analysis of the 2D $\Delta\phi(x,\tau)$ images, $\Delta\Phi(x,\tau)$ is extracted using standard interferogram analysis techniques [19, 20], and $I_S(x,\tau)$ is extracted using a low pass image filter (suppressing the sideband imposed by the transient grating). While we can extract 2D amplitude and phase maps from a single shot, averaging shots yields a better signal-to-noise ratio. However, due to mechanical vibrations in the lab, the fringe positions are not stable shot-to-shot. To prepare frames for averaging, the phase of the fringes in each frame is shifted to enforce alignment. The frames are then averaged, after which the peak of the Fourier side band is windowed and shifted to zero in the frequency domain.

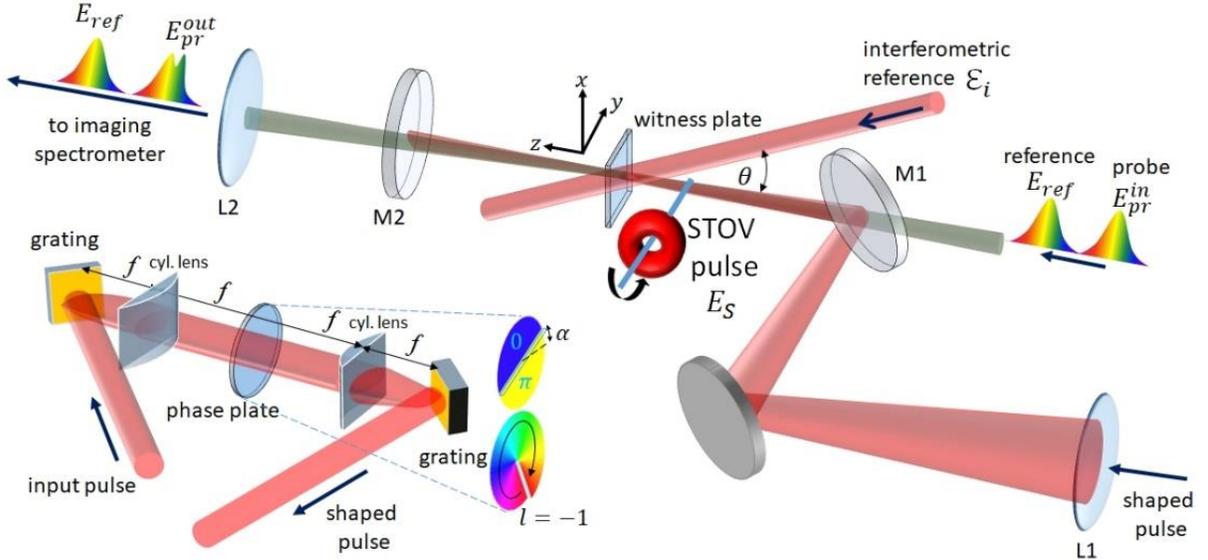

Fig 1. Top: Setup for transient grating single-shot spectral interferometry (TG-SSSI). The STOV-carrying pump pulse (center wavelength $\lambda_0 = 800$nm) at the output of a $4f$ pulse shaper is focused (~1.5 µJ) or imaged (~20 µJ) into a 500 µm thick fused silica witness plate. The pump pulse energy is kept sufficiently low so that the STOV pulse propagates nearly linearly in the plate. A probe pulse $\mathcal{E}_i$ ($\lambda_0 = 795$nm, 2 nm bandwidth) crosses the STOV pulse direction at angle $\theta = 6°$, forming a transient grating with modulations $\propto cos(kx\,sin\,\theta + \Delta\Phi(x,\tau))$, where the symbols are defined in the main text and reference coordinates are shown next to the witness plate. The transient grating is probed by SSSI [19,20], which uses ~1.5 ps long chirped supercontinuum reference and probe pulses $E_{ref}$ and $E_{pr}$ ($\lambda_0 \sim 575$ nm). The result is single-shot time and space resolved images of amplitude and phase of STOV-carrying pulses. Bottom left: Cylindrical lens-based $4f$ pulse shaper [16, 17] for imposing a line-STOV on a 45 fs, λ=800nm input pulse. A phase mask is inserted in the Fourier plane at the common focus of the cylindrical lenses. For the current experiment, we use spiral phase masks ($l = 1$, $l = -1$, and $l = 8$) and a $\pi$−step mask, all etched on fused silica, where the $\pi$−step angle $\alpha$ and the spiral orientation (for $l = -1$) are also shown. Both the $l = 1$ and $l = 8$ plates have 16 levels (steps) every $2\pi$. Shaper gratings: 1200 line/mm, cylindrical lenses: focal length 20cm.

STOVs were generated by a cylindrical lens-based $4f$ pulse shaper [16] depicted in the lower left of Fig. 1. The pulse shaper imposes a line-STOV on an input Gaussian pulse (50fs, 1.5-20µJ) using a $2\pi l$ spiral transmissive phase plate (with $l = +1, -1,$ or 8) or a $\pi$-step plate at the shaper's Fourier plane (common focus of the cylindrical lenses). The vertical and horizontal axes on the phase masks lie in the spatial ($x$) and spectral ($\omega$) domains. The phase plate orientations are shown in the figure, where for the step plate, the adjustable angle $\alpha$ is with respect to the spectral (dispersion) direction. While the shaper imposes a spatio-spectral ($x, \omega$) phase at the phase plate, leading to a spatio-temporal ($x, \tau$) pulse immediately at its output at the exit grating (near field), our desired spatial effects appear in the far-field of the shaper, where the desired STOV-carrying pulse emerges. Here, we project to the far-field by focusing the shaper output with lens L1 into the 500µm fused silica witness plate, whereupon it is measured using TG-SSSI. In this context, the subscript on $E_S$ can now be read as referring to a STOV-carrying pulse.

In Fig. 2, row (a) shows the pulse with no phase plate in the pulse shaper. This is the far field output of the shaper as measured by TG-SSSI in the witness plate. The temporal leading edge is at $\tau < 0$. The left column shows $\Delta\phi(x,\tau)$. The fringes are removed with a low pass filter, yielding $I_S(x,\tau)$ in the next column, while in the third column, a high pass filter leaves the fringe image $f(x,\tau)$. The far right column shows the extracted spatio-temporal phase $\Delta\Phi(x,\tau)$. It is seen that the pulse envelope $I_S$ closely agrees with the 50 fs pulse input to the shaper, and that $\Delta\Phi(x,\tau)$ is weakly parabolic in time (small chirp) and relatively flat in space. The slight curvature of the fringes of $f(x,\tau)$ seen in Fig. 2 is attributed to spectral phase mismatch between $E_S$ and $\mathcal{E}_i$.

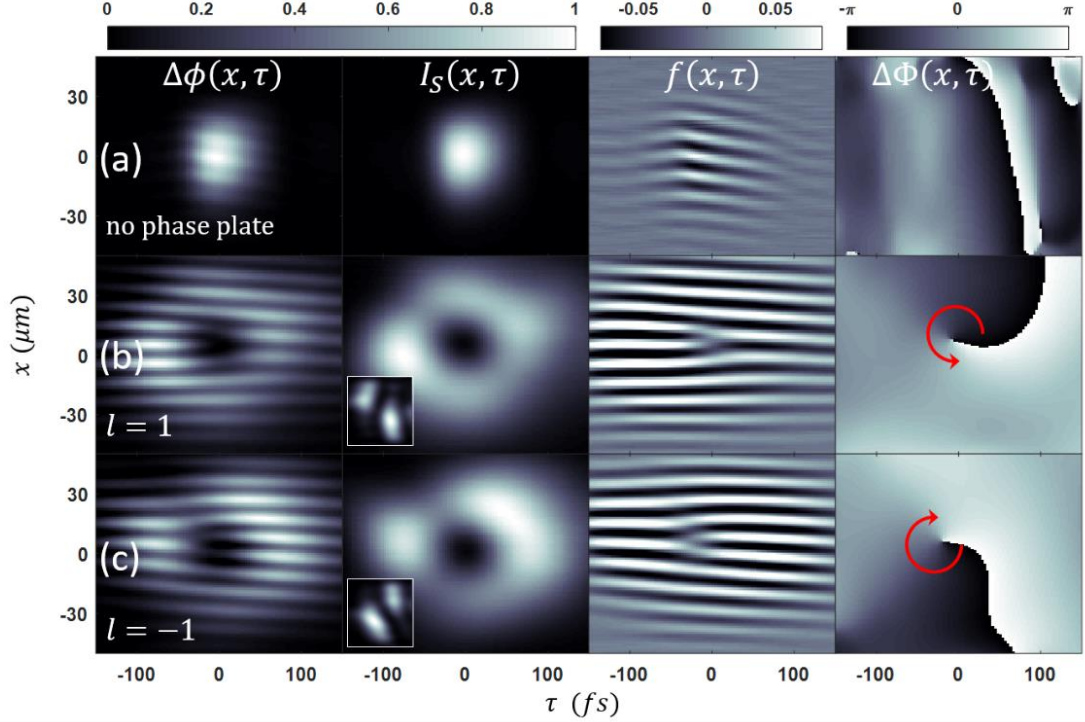

Fig 2. (a) Output of pulse shaper with no phase plate. The 50 fs input pulse, with a weakly parabolic temporal phase, is recovered. (b, c) Intensity and phase of pulse in far-field of pulse shaper with $l=1$ and $l=-1$ spiral phase plates. White-bordered insets: pulse shaper near-field intensity images. The red arrows show the direction of phase circulation. Headings of each column are described in the text. In all panels, the temporal leading edge of the pulse is on the left ($\tau<0$), so propagation is right-to-left. The pulse energy for the three far field cases above is ~1μJ. For the near field cases (insets), the pulse energy is increased to ~20μJ to offset the reduced signal due to magnification.

One form of line-STOV-carrying pulse can be generated with a spiral phase plate in the pulse shaper. For a $l=1$ plate, row (b) of Fig. 2 shows, as in (a), the various extractions from TG-SSSI. The presence of a spatio-temporal phase singularity is evident from the characteristic forked pattern in $f(x,\tau)$. The spatio-temporal envelope $I_S(x,\tau)$ and phase $\Delta\Phi(x,\tau)$ of the STOV are shown in the second and fourth columns, where the pulse appears as an edge-first flying donut with a $2\pi$ phase circulation around the phase singularity at the donut null. Using a $l=-1$ plate (flipping the $l=1$ plate) generates the opposite spatio-temporal phase circulation, as seen in row (c). The small insets in (b) and (c) show the corresponding near-field intensity envelopes from the shaper (obtained by imaging the shaper output at the witness plate), consisting of 2 lobes separated by a space-time diagonal. Owing to conservation of angular momentum, the associated spatio-temporal phase windings (not shown) are the same as in the far-field.

Line-STOVs of charge $l=\pm1$ can also be generated with a $\pi$-step phase plate in the shaper's Fourier plane, rotated to an angle $\alpha_{step}$ with respect to the grating dispersion direction, so that the step lies along the spatio-spectral line $d\bar{x}/d\bar{\omega} = \mp\frac{1}{2}(x_s/x_0)(\tau_s/\tau_0)^{-1}$ (see discussion below), where $x_0$ and $\tau_0$ are the width and duration of the shaper input pulse, with $\bar{x} = $

$x/x_0$ and $\bar{\omega} = \omega\tau_0$. In practice, $\alpha_{step}$ is finely adjusted to get a line-STOV output as measured by TG-SSSI. As seen in Fig. 3, for $\alpha_{step} = 25°$, the near-field output of the shaper is a flying donut (row (a)) with $l = 1$, while the lens-focused, far-field envelope [row (b)] is two lobes separated by a space-time diagonal. Going to $\alpha_{step} = -25°$ [row (c)] gives a STOV-carrying pulse envelope that is the space reflection of (b). In (b) it is seen that the vortex charge adds to $+1$ [consistent with (a)] and in (c) the charge adds to $-1$.

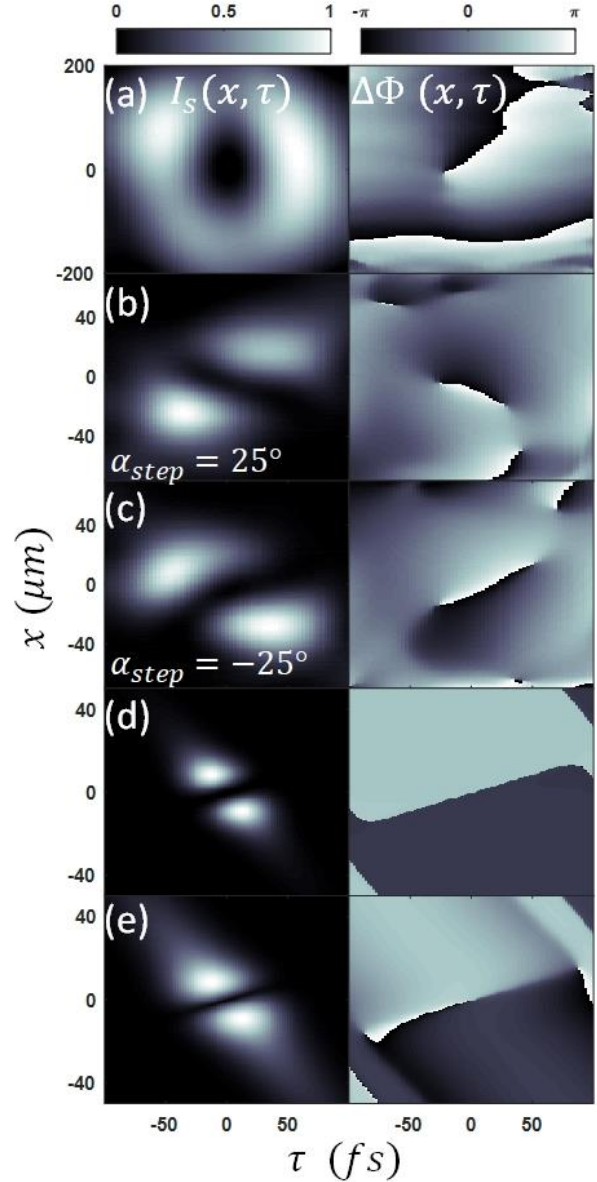

Fig 3. (a) Flying donut near-field intensity and phase from π-step pulse shaper ($\alpha_{step} = +25°, l = 1$), obtained from imaging shaper output into witness plate. (b, c) Offset lobe far-field intensity and phase, obtained by focusing shaper output into witness plate for step orientations $\alpha_{step} = \pm25°$ ($l = \pm1$). (d) Simulation for $\alpha_{step} = -25°$ of far-field intensity and phase for (d) no dispersion, (e) group dispersion delay GDD=100 fs². The addition of parabolic temporal phase to the spatio-temporal phase step of (d) explains the phase pattern in (b, c). Headings of each column are described in the text. The pulse energy in panel (a) is ~20μJ and ~1μJ in panels (b) and (c). Propagation is right-to-left.

We simulate the near-field output of the pulse shaper by Fourier transforming an input spatio-temporal pulse $E_0(x,\tau)$ to the spatio-spectral domain $\tilde{E}_0(x,\omega)$, applying the spatio-spectral phase shift respresented by the phase mask, along with any dispersion, and then Fourier transforming the field back to the spatio-temporal domain as $E(x,\tau)$. Here we ignore the $y$ dependence, which is near-Gaussian throughout. To simulate the far-field output of the shaper, we transform $E(x,\tau) \to \tilde{E}(k_x,\tau) = E'(x',\tau)$, where $x' \propto k_x$ is the local transverse coordinate in the far field. Simulations of the far-field of the $\pi$-step shaper are shown in panels (d) with no dispersion and (e) with group dispersion delay GDD = 100 fs$^2$, corresponding to the measured $\Delta\Phi(x,\tau)$ in Fig. 2(a). The result of (d) is in agreement with the expression for $\tilde{E}(k_x,y,\tau)$ while (e) resembles the experimental result (b). The origin of this effect is that optimizing the SC pulse for TG-SSSI leaves the pump pulse with a very small chirp (parabolic phase in time, as seen in Fig. 2(a)). Adding this phase to the diagonal $\pi$-step phase of 3(d) gives 3(e). Comparing Figs. 2 and 3, we note that the $\pi$-step and $l = \pm 1$ spiral phase shaper outputs appear to be complementary: the near-field of one these "quasi-modes" corresponds to the far-field of the other. As discussed, going from the Fourier plane in the shaper to the shaper output (near-field) and then to the far-field requires two transforms: $(x,\omega) \to (x,\tau) \to (k_x,\tau)$. If we start with Eq. (1) (for $l = \pm 1$) and ignore $z$, $x \to k_x$ yields $\tilde{E}(k_x,y,\tau) = a(\tau/\tau_s \pm \frac{1}{2}k_x x_0^2/x_s)\tilde{E}_0(k_x,y,\tau)$ and then $\tau \to \omega$ yields $\tilde{E}(k_x,y,\omega) = \frac{1}{2}a(i\omega \tau_0^2/\tau_s \pm k_x x_0^2/x_s)\tilde{E}_0(k_x,y,\omega)$, where we have assumed a pulse shaper input $E_0(\mathbf{r}_\perp,\tau) = \epsilon(y)e^{-(x/x_0)^2}e^{-(\tau/\tau_0)^2}$, with spatial and temporal widths $x_0$ and $\tau_0$, and where $\epsilon(y)$ in our experiment is near-Gaussian ($\propto e^{-(y/y_0)^2}$) but can be arbitrary but bounded. However, we can swap $k_x \leftrightarrow 2ix/x_0^2$ and $\omega \leftrightarrow -2i\tau/\tau_0^2$ in any of these expressions to calculate the field at any location given either of the other two. Therefore, a flying donut STOV with a $l = \pm 1$ spiral phase in $(x,\tau)$ in the far-field requires a $l = \pm 1$ spiral phase plate in $(x,\omega)$ in the shaper. A flying donut in $(x,\tau)$ in the near-field requires a $\pi$-step plate in $(x,\omega)$ in the shaper, which yields spatio-temporally offset lobes in the far-field separated by a $\pi$-step in phase.

To estimate the optimum angle $\alpha_{step}$ for the $\pi$-step plate to produce a near-field $l = \pm 1$ STOV at the shaper output, making the appropriate swap in the above expressions gives $\tilde{E}(x,y,\omega) = ia(\frac{1}{2}\omega \tau_0^2/\tau_s \pm x/x_s)\tilde{E}_0(x,y,\omega)$ at the phase plate, where a $\pi$ phase shift occurs across the line $\frac{1}{2}\omega \tau_0^2/\tau_s \pm x/x_s = 0$. The spatio-spectral orientation of the plate's $\pi$-step is therefore $d\bar{x}/d\bar{\omega} = \mp\frac{1}{2}(x_s/x_0)(\tau_s/\tau_0)^{-1}$, as cited earlier, and clearly enables control of the STOV space-time aspect ratio. For example, we have observed that for $\alpha_{step} \to 0$, the STOV appears as 2 lobes reflected across the time axis. This is consistent with $d\bar{x}/d\bar{\omega} \to 0$ and $\tau_0/\tau_s \to 0$, corresponding to extreme time-axis-stretching of the donut hole.

As most experiments with STOVs will take place in the far-field of a pulse shaper, selecting among a flying donut, spatio-temporally offset lobes, or other possible space-time structures will depend on the far-field STOV profile desired for applications. In any case, the electromagnetic angular momentum is conserved through the spatio-temporal / spatio-spectral domains.

To visualize how a STOV-carrying pulse evolves from the near field at the pulse shaper to the far field, we have performed 3D+time UPPE (unidirectional pulse propagation equation) propagation simulations [23,24], as shown in Fig. 4. The input to the shaper is $E_0(\mathbf{r}_\perp,\tau) = \epsilon_0 e^{-(y/y_0)^2}e^{-(x/x_0)^2}e^{-(\tau/\tau_0)^2}$, to which is applied the $l = +1$ spiral spatio-spectral phase factor $e^{i\Delta\varphi(x,\omega)} = \mathbb{Z}/|\mathbb{Z}|$ (phase-only mask corresponding to our experiment), where $\mathbb{Z} =$

$a(\frac{1}{2}\bar{\omega}(\tau_0/\tau_s) + i\bar{x}(x_0/x_s))$, the prefactor of $\tilde{E}_0(x, y, \omega)$ as calculated using the theoretical treatment above. The pulse was then propagated to the far field through a $3m$ lens at Rayleigh range $z_R = 2.3m$ (our finite GPU-based computer memory limited the simulations to lower spatial resolution, necessitating use of a long focal length lens). The top and bottom row of panels in Fig. 4 show amplitude $I_S(x,\tau)$ and phase $\Delta\Phi(x,\tau)$ of the STOV. The $y$-dependence maintains its Gaussian envelope. The white-bordered insets in the bottom row show simulations with the phase *and* amplitude mask $\mathbb{Z}$ applied, corresponding to our theoretical treatment above, which is based on the form of STOV assumed in Eq. (1). The results for both masks are very similar, and either work to generate STOVs. The simulation clearly shows the continuous evolution of the STOV pulse from space-time diagonally-separated lobes to donut, with the STOV charge (and angular momentum) conserved throughout. It is important to reiterate that while the form of STOV assumed in Eq. (1) necessitates a spatio-spectral phase and amplitude mask of form $\mathbb{Z}$, we actually use a pure phase mask $\mathbb{Z}/|\mathbb{Z}|$ in our experiment—that is, our $(x,\omega)$ pulse profile is mismatched to the phase plate profile—but as shown by our 3D+time propagation simulations, this leads to very similar results.

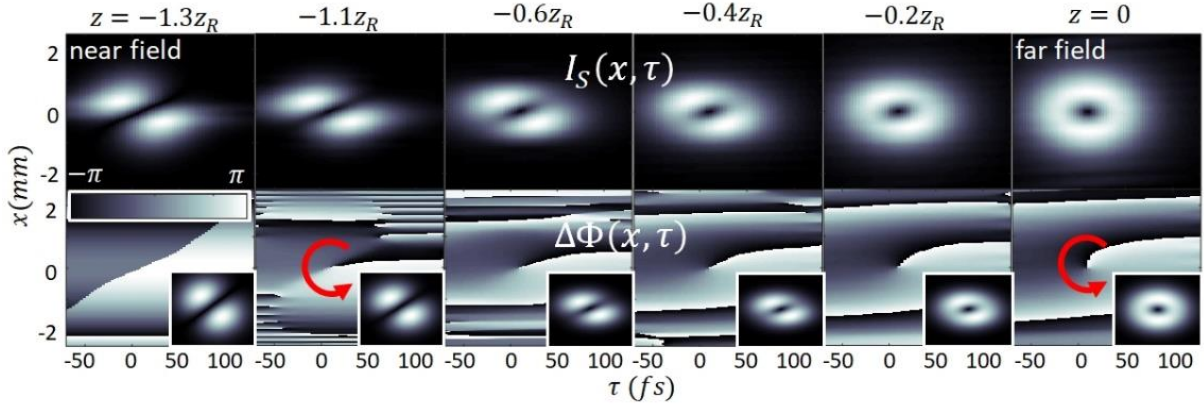

Fig. 4. 3D+time UPPE simulation of STOV-carrying pulse launched from a pulse shaper with a $l = +1$ spatio-spectral spiral phase factor $e^{i\Delta\varphi(x,\omega)} = \mathbb{Z}/|\mathbb{Z}|$ corresponding to our experiment (see text). "Near-field" is right after a $3m$ lens at the pulse shaper output; "far-field" is at the lens focus. Rayleigh range is $z_R = 2.3m$. The propagation direction within each panel is right-to-left. Top row: intensity profiles $I_S(x,\tau)$. Bottom row: phase profiles $\Delta\Phi(x,\tau)$ with red arrows showing phase increase direction. Bottom row white-bordered insets: intensity profiles simulated using $l = +1$ spatio-spectral spiral phase factor $\mathbb{Z}$.

The transformation of one quasi-mode into the other can be viewed as STOV mediation of the energy flow within the pulse. In a frame moving at the pulse group velocity, as shown in [25] and more recently applied to STOVs [12], the local Poynting flux consistent with the paraxial wave equation is $\mathbf{S} = (c/8\pi k_0)|E_S|^2(\nabla_\perp\Phi_{s-t} - \beta_2(\partial\Phi_{s-t}/\partial\xi)\hat{\xi})$, where $\xi = v_g\tau$, $\Phi_{s-t}$ is the spatio-temporal phase (see Eq. (1)), $\hat{\xi}$ is a unit vector along $\xi$, and $\beta_2 = c^2 k_0(\partial^2 k/\partial\omega^2)_0$ is the normalized group velocity dispersion, where $\beta_2^{air} \sim 10^{-5}$ and $\beta_2^{glass} \sim 2 \times 10^{-2}$. Because the first term in $\mathbf{S}$ is dominant for both air and glass, the weakly saddle-shaped energy flow [12] is mostly along $\pm x$, providing the necessary transformation from donut to spatio-temporally offset lobes or back again. This is a remarkable effect: we note that in a STOV-free beam, the term $\nabla_\perp\Phi$ would act on a local spatial phase curvature to transversely direct energy (diffract) to *both* sides of the beam propagation direction (here $\pm x$ and $\pm y$). However, in a $l = 1$ linear STOV whose axis is along $y$ (see Fig 2(b)), $\nabla_\perp\Phi$ points along $(-x)$ in front of the pulse and along $(+x)$ in the back, directing energy density to one side in front of the pulse and the opposite side in the back. This is seen in the transformation of the spatio-temporally offset lobes from the near field (Fig. 2(b) inset image) to the far-field flying donut. Similar dynamics apply to the $l = -1$ STOV of Fig. 2(c), and to the $l = 1$ STOV of Fig. 3 (a) and (b).

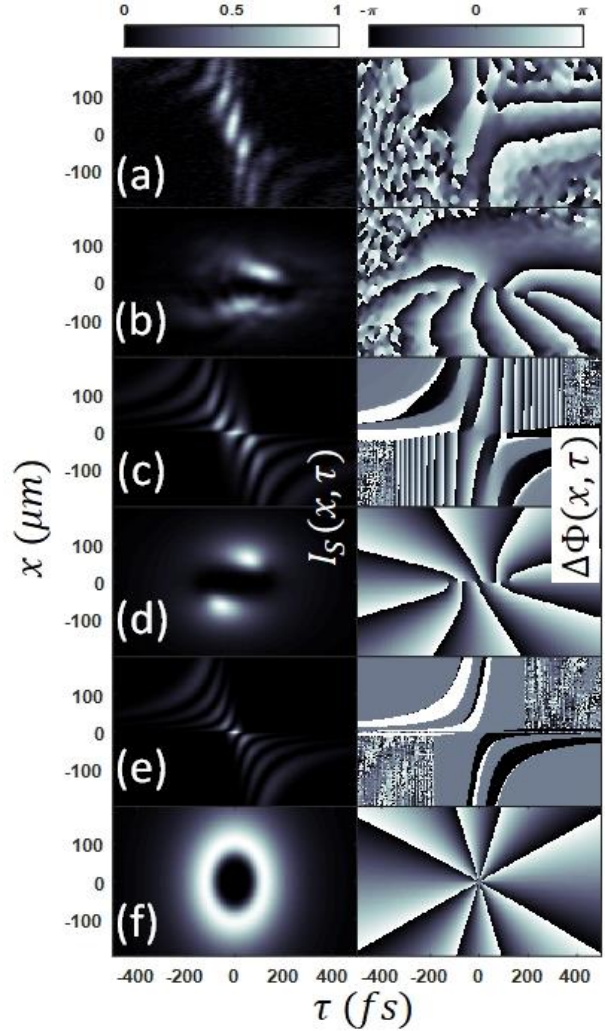

Fig 5. Results from pulse shaper with $l = 8$ spiral phase plate. Pulse propagation is right-to-left. Left column: intensity profiles $I_S(x, \tau)$. Right column: phase profiles $\Delta\Phi(x, \tau)$. All profiles are spatially rescaled for comparison. (a) Near-field intensity and phase of shaper output, obtained from imaging exit grating onto witness plate. The $l = 8$ vortex appears as eight $\pi$-step phase jumps (only 6 visible owing to underfilling of image by probe SC pulse). Laser pulse energy ~20 µJ. (b) Far-field intensity and phase obtained by focusing shaper output into witness plate. Here, eight $l = 1$ STOVs are seen in the phase plot (here, the SC reference pulse profile overfills the smaller spot). Laser pulse energy ~2 µJ. (c,d) Fourier transform simulation of near-field and far-field intensity and phase where the $(x, \omega)$ spatial profile of the pulse in the shaper is *not* matched to the phase mask. (e,f) Fourier transform simulation of near-field and far-field intensity and phase where the $(x, \omega)$ pulse profile in the shaper *is* matched to the phase plate.

To explore higher order STOVs, we use a $l = 8$ ($16\pi$) spiral phase plate in the pulse shaper. Figure 5(a) shows the near-field intensity envelope and phase while 5(b) shows the intensity and phase in the far field. In the near field as shown in (a), six $\pi$-step phase jumps appear, corresponding to nulls in the intensity envelope (rather than 8 because the SC probe pulse underfilled the larger image of the exit grating in the witness plate). In the far field, where the SC probe pulse overfilled the pump pulse, enabling coverage of all the vortices, it is seen that the pulse has formed eight $l = +1$ STOVs. Such splitting of high charge vortices into multiple single charge vortices has been explained for standard monochromatic OAM as originating from interference with a coherent background, or with a coherent probe beam used to measure the presence of vortices [26].

In our case, the splitting has a different origin: a mismatch of our $(x, \omega)$ beam profile in the pulse shaper to the profile of the $l = 8$ spiral phase plate. While this mismatch has only minor effects for generating $l = 1$ STOVs, as discussed in the context of Fig. 4, it reveals itself for higher order STOVs. Fourier transform simulations, including glass dispersion, are shown in Fig. 5(c) and (d) for the near and far fields, reproducing the main features of the measurements, including the "splitting" into eight $l = +1$ vortices. The $(x, \omega)$ beam profile-phase plate mismatch leads to slightly different orientations of adjacent pairs of near-field lobes in Fig. 5(a) (5(c)); these form slightly displaced $l = 1$ windings in the far field in Fig. 5(b) (5(d)). So in our case, it appears that the $l = 8$ STOV never forms and eight $l = 1$ STOVs are formed directly. We expect that careful dispersion management and a better match of our spatio-spectral beam profile with the phase plate will enable generation of high order STOVs that can propagate into the far field. This is shown in the simulations of Fig. 5 (e,f) for the case where the spatio-spectral profile and the phase plate are matched: an $l = 8$ flying donut is formed, accompanied by a single vortex of the same charge.

In conclusion, we have demonstrated the linear generation and propagation in free space of pulses that carry a new type of optical orbital angular momentum whose associated vortex phase circulation exists in space-time: the spatio-temporal optical vortex (STOV). Our measurements show that freely propagating STOVs conserve angular momentum in space-time and mediate space-time energy flow within the pulse. We have introduced a new ultrafast diagnostic, transient grating single-shot supercontinuum spectral interferometry (TG-SSSI) to measure the space- and time-resolved amplitude and phase of a STOV in a single shot. We expect that nonlinear propagation of STOV-carrying pulses or propagation of STOVs through fluctuating media will provide a rich area of study, and in such experiments sensitive to shot-to-shot fluctuations, TG-SSSI will be an important tool.

**Funding.** Air Force Office of Scientific Research (AFOSR) (FA9550-16-10121, FA9550-16-10284); Office of Naval Research (ONR) (N00014-17-1-2705, N00014-17-12778), NSF (PHY1619582).

**Acknowledgements.** The authors thank N. Jhajj and J. Wahlstrand for early work on the pulse shaper, and J. Griff-McMahon and I. Larkin for help with the simulations.

**Disclosures.** The authors declare no conflicts of interest.